\documentclass{ws-p9-75x6-50}
\usepackage{epsfig,pstricks}
\input{psfig.tex}

\newcommand{\gae}{$\stackrel{>}{\sim}$}
\newcommand{\beq}{\begin{displaymath}}
\newcommand{\eeq}{\end{displaymath}}
\newcommand{\beqa}{\begin{eqnarray}}
\newcommand{\eeqa}{\end{eqnarray}}

\def\slashchar#1{\setbox0=\hbox{$#1$}           
   \dimen0=\wd0                                 
   \setbox1=\hbox{/} \dimen1=\wd1               
   \ifdim\dimen0>\dimen1                        
      \rlap{\hbox to \dimen0{\hfil/\hfil}}      
      #1                                        
   \else                                        
      \rlap{\hbox to \dimen1{\hfil$#1$\hfil}}   
      /                                         
   \fi}                                         %

\newcommand{\mev}{{\rm \, MeV}}
\newcommand{\gev}{{\rm \, GeV}}
\newcommand{\tev}{{\rm \, TeV}}

\newcmykcolor{purp}{.2 1 0 .4}
\newcmykcolor{grnn}{1 .2 .5 .4}
\newcmykcolor{mag}{0 1 0 0}
\newcmykcolor{orngg}{.1 .1 1 .2}
\newrgbcolor{brwn}{.5 .35 .1}

\begin{document}

\title{Avenues for Dynamical Symmetry Breaking}

\author{R. Sekhar Chivukula}

\address{Physics Department, Boston University
\\590 Commonwealth Ave., Boston, MA 02215 USA
\\E-mail: sekhar@bu.edu
\\{\tt BUHEP-99-6}}


\maketitle

\abstracts{In this talk I review modern theories of dynamical
  electroweak symmetry breaking and some of their signatures at Run II
  of the Tevatron collider.}

\section{Dynamical Electroweak Symmetry Breaking}

\subsection{Technicolor}

The simplest theory of dynamical electroweak symmetry
breaking is technicolor.\cite{Weinberg:1979bn,Susskind:1978ms} Consider
an $SU(N_{TC})$ gauge theory with fermions in the fundamental
representation of the gauge group
\beq
\Psi_L=\left(
\begin{array}{c}
U\\D
\end{array}
\right) _L\,\,\,\,\,\,\,\,
U_R,D_R~.
\eeq
The fermion kinetic energy terms
for this theory are
\beqa
{\cal L} &=& \bar{U}_L i\slashchar{D} U_L+
\bar{U}_R i\slashchar{D} U_R+\\
 & &\bar{D}_L i\slashchar{D} D_L+
\bar{D}_R i\slashchar{D} D_R~,
\nonumber
\eeqa
and, like QCD in the $m_u$, $m_d \to 0$ limit, they have
a chiral $SU(2)_L \times SU(2)_R$ symmetry.

As in QCD the exchange of technigluons in the spin zero, isospin zero
channel is attractive, causing the formation of a condensate
\beq
{\lower15pt\hbox{
\epsfysize=0.5 truein \epsfbox{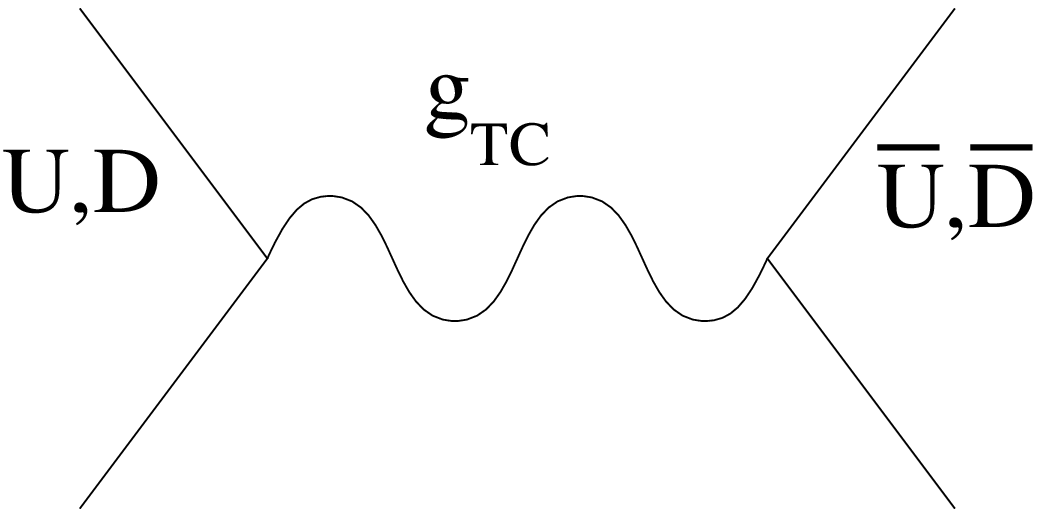}}
\ \ \rightarrow  \ \ \langle \bar U_LU_R\rangle
=\langle \bar D_LD_R\rangle \neq 0\, ,}
\eeq
which dynamically breaks $SU(2)_L \times SU(2)_R \to SU(2)_V$.  These
broken chiral symmetries imply the existence of three massless Goldstone
bosons, the analogs of the pions in QCD.

Now consider gauging $SU(2)_W \times U(1)_Y$ with the left-handed
fermions transforming as weak doublets and the right-handed ones as weak
singlets. To avoid gauge anomalies, in this one-doublet technicolor
model we will take the left-handed technifermions to have hypercharge
zero and the right-handed up- and down-technifermions to have
hypercharge $\pm 1/2$.  The spontaneous breaking of the chiral symmetry
breaks the weak-interactions down to electromagnetism. The would-be
Goldstone bosons become the longitudinal components of the $W$ and $Z$
\beq
\pi^\pm,\, \pi^0 \, \rightarrow\, W^\pm_L,\, Z_L~,
\eeq
which
acquire a mass
\beq
M_W = {g F_{TC} \over 2}~.
\eeq
Here $F_{TC}$ is the analog of $f_\pi$ in QCD. In order
to obtain the experimentally observed masses, we must have
that $F_{TC} \approx 246\, {\rm GeV}$ and hence this model
is essentially QCD scaled up by a factor of
\beq
{F_{TC}\over f_\pi} \approx 2500\, .
\eeq
As in QCD, we expect resonances analogous to the vector mesons 
(the $\rho$ and $\omega$) followed by a tower of higher mass states.

\subsection{Fermion Masses \& ETC Interactions}

In order to give rise to masses for the ordinary quarks and leptons, we
must introduce interactions which connect the chiral-symmetries of
technifermions to those of the ordinary fermions. The most popular
choice\cite{Eichten:1979ah,Dimopoulos:1979es} is to embed technicolor
in a larger gauge group, called {\it extended technicolor} (ETC), which
includes both the unbroken technicolor interactions plus new massive
gauge bosons which couple technifermions ($T$) to ordinary fermions ($f$):
\beq
\hspace{.25cm}{\lower15pt\hbox{\epsfysize=.25 truein
      \epsfbox{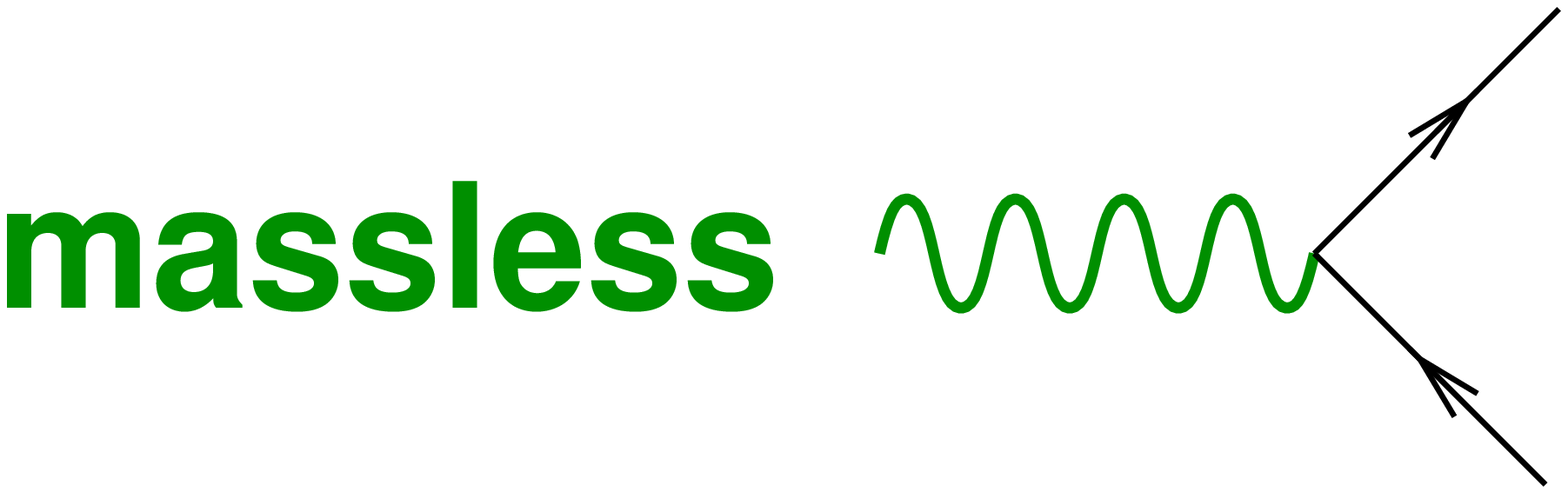}}}\hspace{.25cm}
\hbox{$  \left(\matrix{\bf {\blue f}\cr{\bf T}
    \cr { \bf T}\cr}\right)$ } 
\hspace{.10cm} {\raise1pt\hbox{\epsfysize=.25 truein
      \epsfbox{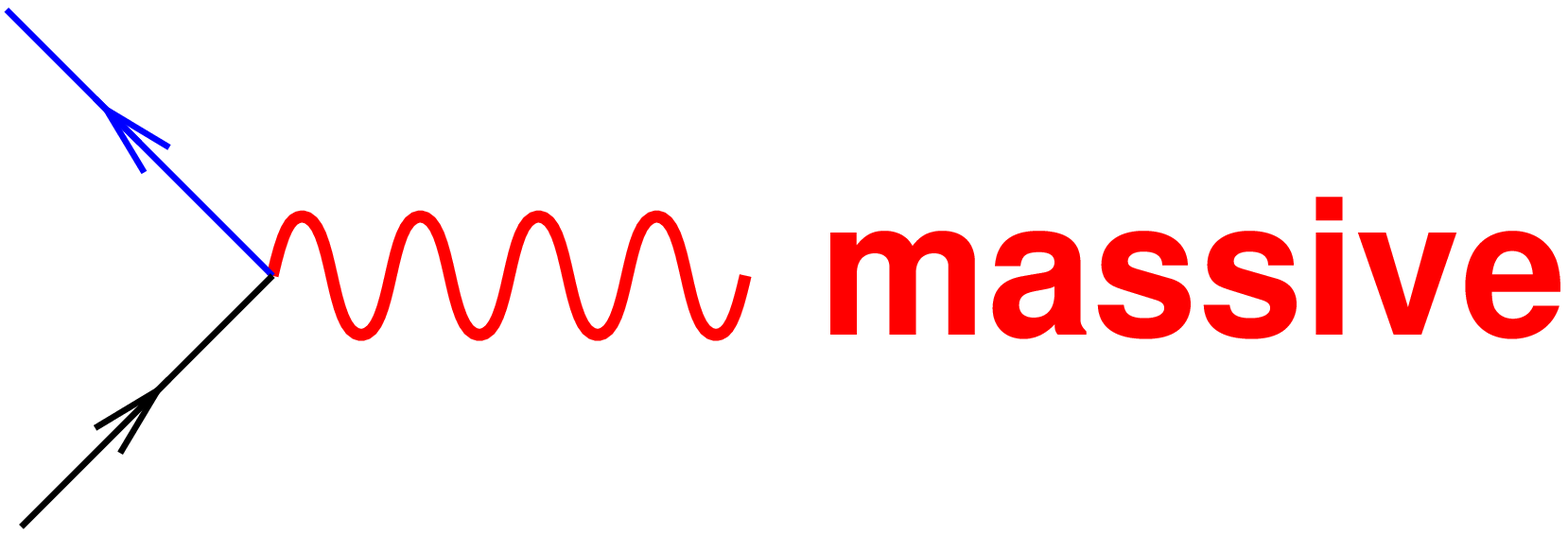}}}
\eeq

At energies low compared to the ETC gauge-boson mass, $M_{ETC}$, these
effects can be treated as local four-fermion interactions
\beq
{\lower15pt\hbox{\epsfysize=0.5 truein \epsfbox{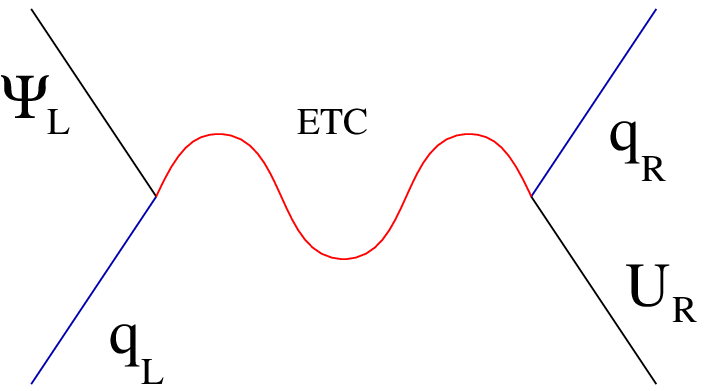}}}
\ \ \rightarrow\ \  {{g_{ETC}^2\over M^2_{ETC}}}(\overline{\Psi}_L U_R)
({\overline{q}_R q_L})~.
\label{etcint}
\eeq
After technicolor chiral-symmetry breaking and the formation of a
$\langle \bar{U} U \rangle$ condensate, such an interaction gives rise
to a mass for an ordinary fermion
\beq
m_q \approx {{g_{ETC}^2\over M^2_{ETC}}} \langle\overline{U} U\rangle_{ETC}~,
\label{fmass}
\eeq
where $\langle \overline{U} U\rangle_{ETC}$ is the value of the
technifermion condensate evaluated at the ETC scale (of order
$M_{ETC}$).  The condensate renormalized at the ETC scale in 
can be related to the condensate renormalized at the
technicolor scale as follows
\beq
\langle\overline{U} U\rangle_{ETC} = \langle\overline{U} U\rangle_{TC}
\exp\left(\int_{\Lambda_{TC}}^{M_{ETC}} {d\mu \over \mu}
\gamma_m(\mu)\right)~,
\eeq
where $\gamma_m(\mu)$ is the anomalous dimension of the
fermion mass operator and $\Lambda_{TC}$ is the analog of $\Lambda_{QCD}$
for the technicolor interactions.

For QCD-like technicolor (or any theory which is ``precociously''
asymptotically free), $\gamma_m$ is small over in the range between
$\Lambda_{TC}$ and $M_{ETC}$ and using dimensional
analysis\cite{Weinberg:1979kz,Georgi1984,Manohar:1984md,Gasser:1984yg,Gasser:1985gg}
we find
\beq
\langle\overline{U} U\rangle_{ETC} \approx \langle\overline{U} U\rangle_{TC}
\approx 4\pi F^3_{TC}~.
\eeq
In this case we find that
\beq
{{M_{ETC}\over g_{ETC}}} \approx 40 \tev 
\left({F_{TC}\over 250\gev}\right)^{3\over 2}
\left({100 \mev \over m_q}\right)^{1\over 2}~.
\eeq

\subsection{Flavor-Changing Neutral-Currents}

Perhaps the single biggest obstacle to constructing a realistic ETC
model is the potential for
flavor-changing neutral currents.\cite{Eichten:1979ah}  Quark mixing implies
transitions between different generations: $q \to \Psi \to q^\prime$,
where $q$ and $q'$ are quarks of the same charge from different
generations and $\Psi$ is a technifermion. Consider the commutator of
two ETC gauge currents:
\beq
[\overline{q}\gamma \Psi, \overline{\Psi}\gamma q^\prime]\, \supset\, 
\overline{q}\gamma q^\prime\, .
\eeq
Hence we expect there are ETC gauge bosons which couple to {\it
  flavor-changing neutral currents}. In fact, this argument is slightly
too slick: the same applies to the charged-current weak interactions!
However in that case the gauge interactions, $SU(2)_W$ respect a global
$(SU(5) \times U(1))^5$ chiral symmetry \cite{Chivukula:1987py} leading
to the usual {GIM} mechanism.\cite{Glashow:1970st}

Unfortunately, the ETC interactions {cannot} respect GIM exactly;
they must distinguish between the various generations in order to give
rise to the masses of the different generations. Therefore, flavor-changing
neutral-current interactions are (at least at some level) unavoidable.

The most severe constraints come from possible $|\Delta S| = 2$
interactions which contribute to the $K_L$-$K_S$ mass
difference. In particular, we would expect that in order to produce
Cabibbo-mixing the same interactions which give rise to the $s$-quark
mass could cause the flavor-changing interactions
which contribute to the neutral kaon mass splitting.
Experimentally we know that
$\Delta M_K <   3.5 \times 10^{-12}\,\mev $. We then calculate, 
using the vacuum insertion approximation, that
\beq
{M_{ETC} \over {g_{ETC} \, \sqrt{{\rm Re}(\theta^2_{sd})}}} >  600\,\tev~,
\eeq
where $\theta_{sd}$ is of order the Cabibbo angle.
Using the relation of the ETC gauge-boson mass to the fermion mass, we find that
\beq
m_{q, \ell} \simeq {g_{ETC}^2 \over {M_{ETC}^2}}
\langle\overline{T}T\rangle_{ETC}  < {0.5\,\mev\over{N_D^{3/2} \, \theta_{sd}^2}} \,
\eeq
showing that it will be difficult to produce the {$s$}-quark
mass, let alone the {$c$}-quark!

\subsection{Walking Technicolor}

We must therefore conclude that, to be viable, {\it technicolor dynamics
  cannot be like QCD}\,! How could it be different? Recall that the
estimates given above result from the assumption that $\gamma_m \simeq
0$, {\it i.e.} that the theory is precociously asymptotically free. On
the other hand, if $\beta(\alpha_{TC}) \simeq 0$ all the way from
$\Lambda_{TC}$ to $M_{ETC}$, then the technicolor coupling remains {\it
  strong} and close to the value required to produce chiral symmetry
breaking. That is, the coupling ``walks'' between the technicolor and
extended technicolor scales. If this is the case, it is believed
that\cite{Holdom:1981rm,Holdom:1985sk,Yamawaki:1986zg,Appelquist:1986an,Appelquist:1987tr,Appelquist:1987fc}
$\gamma_m(\mu) \cong 1$ in this range. We then find 
\beq
m_{q,l} = {g^2_{ETC} \over {M^2_{ETC}}} \times
\left(\langle\overline{T}T\rangle_{ETC} \cong 
\langle\overline{T}T\rangle_{TC} \, {M_{ETC} \over {\Lambda_{TC}}} \right)~.
\eeq

We have previously estimated that flavor-changing
neutral current requirements imply that the
ETC scale associated with the second generation must
be greater than of order 100 to 1000 TeV. The walking
technicolor enhancement of the technifermion condensate implies that
\beq
m_{q,l} \simeq {{50\, -\, 500\mev}\over N^{3/2}_D \theta^2_{sd}}~,
\eeq
arguably enough to accommodate the strange and charm quarks.

\section{Low-Scale Technicolor at the Tevatron}

How can $\beta(\alpha_{TC}) \simeq 0$? The gauge contributions to
the $\beta$ function are {\it always} negative; therefore these
must be canceled by fermions. To cancel the gauge contribution
entirely one must introduce either {\it many fermions}, or
fermions in {\it higher representations} of the gauge group (or possibly
both). 

Increasing the number of fermions enlarges the chiral symmetries which
are present.  If the chiral symmetry is larger than $SU(2)_L\times
SU(2)_R$, there will be additional (pseudo-)Goldstone bosons ($\pi_T$)
which are not ``eaten'' by the $W$ and $Z$. In general, such nonminimal
models will contain several sets of electroweak doublets. The
$F$-constant associated with each sector is analogous to the vacuum
expectation values $v_i$ of the different Higgs scalars in a multi-Higgs
model.  The lower the value of the $F$-constant, the lower the masses of
the corresponding resonances.

\begin{figure}[tbp]
\begin{center}
\epsfxsize=6cm
\epsffile{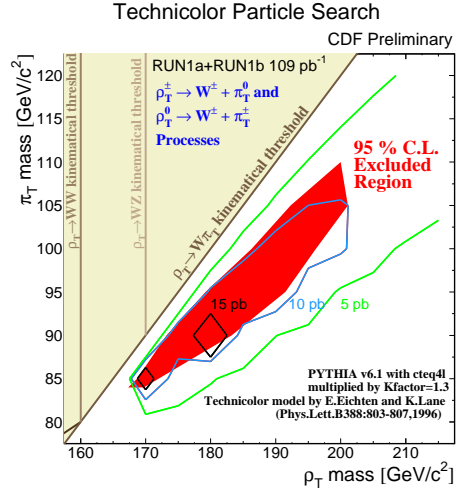}
\end{center}
\caption{95\% Exclusion region\protect\cite{HANDA} for light technirho's decaying
to $W^\pm$ and a $\pi_T$, and in which the $\pi_T$ decays
to two jets including at least one $b$-quark.} 
\label{Fig1}
\end{figure}

These considerations lead to the possibility of discovering signals of a
low-scale technicolor sector at the Tevatron.\cite{Eichten:1996dx} The
extended symmetry breaking sector could give rise to potentially light
resonances, such as a technirho in the few-100 GeV range. It might be
expected, in analogy with QCD, that such a technirho would decay
dominantly to technipions. However, in walking technicolor the effects
of ``small'' chiral-symmetry breaking interactions are likely to be
enhanced. It is then possible that $\rho_T \to \pi_T \pi_T$ is {\it
  closed}.  In this case the dominant decay mode is $\rho_T \to W_L
\pi_T$, and the technirho can be very narrow. As in the case of extra
Higgs scalars in multihiggs models, we expect the technipions to decay
to the heaviest fermions, $\pi^0_T \to b\bar{b}$ \& $\pi^\pm_T \to
c\bar{b}$. Hence, we consider the overall signal $\rho^\pm_T \to W^\pm
\pi^0_T \to \ell \nu (j j)\vert_b$:
\beq
\hbox{\epsfxsize 3cm \centerline{\epsffile{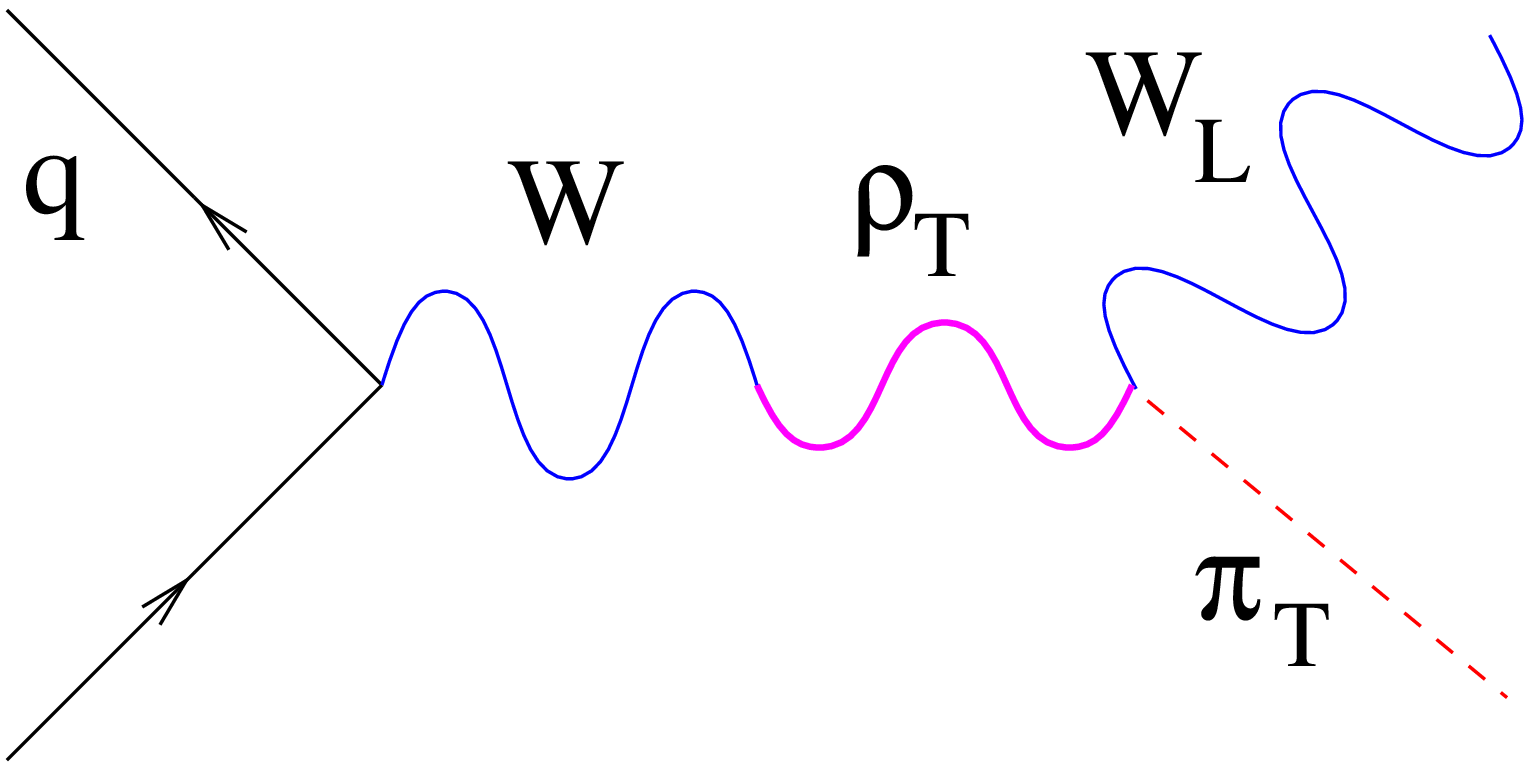}}}
\eeq
Simulations\cite{Eichten:1997yq} at the Tevatron yield a potentially
observable cross section: $\sigma_{\rho_T} \cdot {\rm BR}$ = {5.3 pb}.

Recently, a search has been done in this channel by the CDF
collaboration\cite{HANDA} based on Run I data. These results are shown
in figure \ref{Fig1}, and we see that $\sigma \cdot {\rm BR} \geq 15$ pb
is already excluded at the 95\% confidence level. We expect in Run II,
with twenty times as much data, the sensitivity will reach the predicted
level.

\section{Topcolor-Assisted Technicolor (TC2)}

The top-quark is much heavier than other fermions.  It must therefore be
more strongly coupled to the symmetry-breaking sector.  Perhaps all or
some of electroweak-symmetry breaking is due to a condensate of
top-quarks,\cite{Miranskii:1989ds,Miranskii:1989xi,Nambu:1989jt,Marciano:1989xd,Bardeen:1990ds,Hill:1991at,Cvetic:1997eb}
$\langle \bar{t}t\rangle \neq 0$.

Recently, Chris Hill has proposed\cite{Hill:1995hp} a theory which
combines technicolor and top-condensation. Features of this type of
model include technicolor dynamics at 1 TeV, which dynamically generates
{most} of electroweak symmetry breaking, and extended technicolor
dynamics at scales much higher than 1 TeV, which generates the light
quark and lepton masses, as well as small contributions to the third
generation masses ($m_{t,b,\tau}^{ETC}$) of order 1 GeV. The top quark
mass arises predominantly from ``topcolor,'' a new QCD-like
interaction which couples preferentially to the third generation of
quarks, at a scale of order 1 TeV and which generates $\langle \bar{t} t
\rangle \neq 0$ and $m_t \sim 175$ GeV. 

\medskip

{\underline{Hill's Simplest TC2 Scheme}}

\medskip

The simplest scheme\cite{Hill:1995hp} which realizes these features has the
following structure:
\medskip
\begin{center}
$G_{TC}  \times  SU(2)_{EW} \times 
SU(3)_{tc} \times SU(3) \times U(1)_H \times U(1)_L$
\end{center}
\smallskip
\begin{center}
$\downarrow \ \ M$ \gae 1 TeV
\end{center}
\smallskip
\begin{center}
$G_{TC} \times SU(3)_C  \times SU(2)_{EW} \times U(1)_Y $
\end{center}
\smallskip
\begin{center}
$\downarrow\ \ \ \Lambda_{TC}\sim 1{\ \rm TeV} $
\end{center}
\smallskip
\begin{center}
$SU(3)_C \times U(1)_{EM}$
\end{center}
\medskip
Here $U(1)_H$ and $U(1)_L$ are $U(1)$ gauge groups coupled to the
(standard model) hypercharges of the third-generation and first-two generation
fermions respectively.
Below $M$, this leads to the effective interactions:
\beq
-{{4\pi \kappa_{tc}}\over{M^2}}\left[\overline{\psi}\gamma_\mu 
{{\lambda^a}\over{2}} \psi \right]^2~,
\eeq
from topgluon exchange and the isospin-violating interactions
\beq
-{{4\pi \kappa_1}\over{M^2}}\left[{1\over3}\overline{\psi_L}\gamma_\mu  \psi_L
+{4\over3}\overline{t_R}\gamma_\mu  t_R
-{2\over3}\overline{b_R}\gamma_\mu  b_R
\right]^2\, ,
\label{hyper}
\eeq
from exchange of the ``heavy-hypercharge'' ($Z^\prime$) 
gauge boson.

The interactions above are attractive in the $\bar{t}t$ channel, but
repulsive in the $\bar{b}b$ channel and the couplings $\kappa_{tc}$ and
$\kappa_1$ can be chosen to produce $\langle \bar{t} t \rangle \neq 0$
and a large $m_t$, but leave $\langle \bar{b} b \rangle = 0$.

\section{Topgluon Searches at the Tevatron}

\begin{figure}[tbp]
\begin{center}
\epsfxsize=8cm
\centerline{\epsfbox{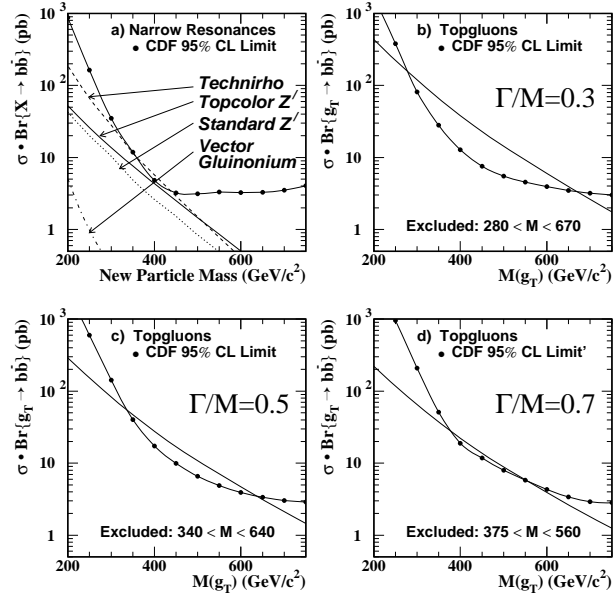}}
\end{center}
\caption{Topgluon searches at
the Tevatron\protect\cite{Abe:1998uz}
in the mode $p{\bar p} \to { g_{TC}} + X \to { b\bar{b}} + X$.}
\label{Figa}
\end{figure}

\begin{figure}[tbp]
\begin{center}
\epsfxsize=8cm
\centerline{\epsfbox{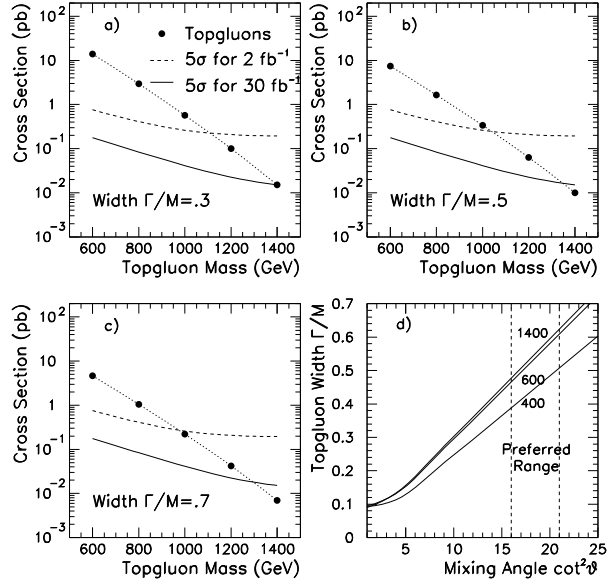}}
\end{center}
\caption{Projected\protect\cite{Harris:1996cu} topgluon reach
at the Tevatron in the mode $p{\bar p} \to { g_{TC}} + X \to { t\bar{t}} + X$.}
\label{Figb}
\end{figure}

The topgluon is a massive color-octet vector which couples
preferentially to the third generation. It has been
searched\cite{Abe:1998uz} for by CDF in the mode $p{\bar p} \to {g_{TC}}
+ X \to {b\bar{b}} + X$.  The results are shown in figure
\ref{Figa}. As shown, topgluon masses from approximately 300 to 600 GeV
are excluded at 95\% confidence level, depending on the width
of the topgluon. Furthermore, as shown in figure \ref{Figb} for the mode
$g_{TC} \rightarrow t\bar{t}$, the Tevatron\cite{Harris:1996cu} should
be sensitive to topgluons up to masses of order 1 TeV in Run II.

\section{All Symmetry Breaking is Dynamical...}

The models presented above may seem rather complicated, especially in
contrast to the one Higgs doublet standard model. However, the the standard
model itself must have some underlying dynamics which gives rise to
symmetry breaking!  In quantum field theory, the vacuum acts as a
polarizable medium: all couplings are a function of the momentum scale
at which they are measured. In the symmetry breaking sector of the
standard one-doublet higgs model, the relevant coupling is the
self-coupling of the Higgs boson.\footnote{For convenience, we ignore
  the corrections due to the weak gauge interactions.  In perturbation
  theory, at least, the presence of these interactions does not
  qualitatively change the features of the Higgs sector.} The dependence
of this coupling as a function of momentum scale is determined by the
$\beta$ function. To lowest-order in perturbation theory, we find
\beq
{\lower7pt\hbox{\epsfysize=0.25 truein \epsfbox{figures/beta.eps}}}
\ \rightarrow \ \beta = {3\lambda^2 \over 2 \pi^2} \, > \, 0
{}~.
\eeq

Integrating this positive $\beta$ function, we find that the effective
coupling becomes infinite at a {\it finite} momentum scale. Conversely
if we require the theory make sense to arbitrarily high momentum scales,
{\it i.e.} that it is truly fundamental, the renormalized coupling at
low-energies is zero!  That is, of we try to take the continuum limit
the theory becomes free (and is hence
trivial\cite{Wilson:1971dh,Wilson:1974dg}), and could not result in the
observed symmetry breaking. The argument given here relies on
perturbation theory, but non-perturbative
investigations\cite{Kuti:1988nr,Luscher:1989uq,Hasenfratz:1987eh,Hasenfratz:1989kr,Bhanot:1990zd,Bhanot:1991ai}
confirm the triviality of the standard higgs model.

The triviality of the scalar sector of the standard one-doublet Higgs
model implies that this theory is only an effective low-energy theory
valid below some cut-off scale $\Lambda$.  Physically this scale marks
the appearance of new strongly-interacting symmetry-breaking dynamics.
Examples of such high-energy theories include ``top-mode'' standard
models\cite{Miranskii:1989ds,Miranskii:1989xi,Nambu:1989jt,Marciano:1989xd,Bardeen:1990ds,Hill:1991at,Cvetic:1997eb}
and composite Higgs
models.\cite{Kaplan:1984fs,Kaplan:1984sm,Dugan:1985hq}

In this sense, {\it all symmetry breaking is dynamical}!  A {light}
Higgs boson, such as occurs in the minimal supersymmetric model, arises
from a dynamical theory at { high $\Lambda$}. However, a {\it heavy}
Higgs requires a large low-energy self-coupling, and therefore arises
from an underlying strongly-interacting theory at relatively low
$\Lambda$.

\section{Conclusions}

Technicolor, topcolor, and related models provide an avenue for
constructing theories in which electroweak symmetry breaking is natural
and has a dynamical origin. Unfortunately, no complete and consistent
model of this type exist.  This is not surprising, since such a theory
must also be a theory of flavor. If electroweak symmetry breaking is due
to strong dynamics at energy scales of order a TeV, experimental
direction will be crucial to construct the correct theory.  With luck,
the necessary clues will begin to appear at the Tevatron in Run II!

\section*{Acknowledgments}

I would like to express my gratitude to V.~S.~Narasimham, Naba Mondal,
and the rest of the conference organizers for a most enjoyable meeting!

{\em This work was supported in part by the Department of Energy under
  grant DE-FG02-91ER40676.}


\end{document}